\documentclass{Interspeech}

 \interspeechcameraready

\title{Zero-Shot Speech LLMs for Multi-Aspect Evaluation of L2 Speech: Challenges and Opportunities}

\author{Aditya Kamlesh}{Parikh}
\author{Cristian}{Tejedor-Garcia}
\author{Catia}{Cucchiarini}
\author{Helmer}{Strik}

\affiliation{Centre for Language Studies}{Radboud University}{the Netherlands}

\email{aditya.parikh@ru.nl, cristian.tejedorgarcia@ru.nl, catia.cucchiarini@ru.nl, helmer.strik@ru.nl}

\usepackage{comment}
\usepackage{float}
\usepackage[utf8]{inputenc}
\usepackage{listings}
\usepackage{xcolor}

\begin{document}

\maketitle

\begin{abstract}
An accurate assessment of L2 English pronunciation is crucial for language learning, as it provides personalized feedback and ensures a fair evaluation of individual progress. However, automated scoring remains challenging due to the complexity of sentence-level fluency, prosody, and completeness. This paper evaluates the zero-shot performance of Qwen2-Audio-7B-Instruct, an instruction-tuned speech-LLM, on 5,000 Speechocean762 utterances. The model generates rubric-aligned scores for accuracy, fluency, prosody, and completeness, showing strong agreement with human ratings within ±2 tolerance, especially for high-quality speech. However, it tends to overpredict low-quality speech scores and lacks precision in error detection. These findings demonstrate the strong potential of speech LLMs in scalable pronunciation assessment and suggest future improvements through enhanced prompting, calibration, and phonetic integration to advance Computer-Assisted Pronunciation Training.

    
\end{abstract}

\section{Introduction}

Globalization has increased cross-border movement for work, education, and other opportunities, making local language acquisition vital for integration, career growth, and well-being \cite{zalli2024globalization}. Beyond vocabulary and grammar, clear pronunciation is crucial for intelligibility, confidence, and meaningful interaction \cite{zielinski2012social, jenkins2000phonology}. Poor pronunciation can hinder communication and create barriers in academic and professional contexts \cite{zossadult}. However, the pronunciation of second-language (L2) is difficult to master due to differences from the mother tongue (L1) and a lack of tools for personalized feedback.

Computer-Assisted Language Learning (CALL), particularly Computer-Assisted Pronunciation Training (CAPT), has become essential in language education in the past years \cite{tejedor2020design}. These tools offer scalable, cost-effective, and consistent pronunciation feedback, addressing the limitations of human evaluation. Traditional CAPT systems use phoneme-level scoring methods like Goodness of Pronunciation (GOP), which assess individual sounds for correctness \cite{witt2000phone}. However, they often fail to address sentence-level fluency, prosody, and completeness, and typically do not provide corrective feedback to learners \cite{gong2022transformer}.

Recent advancements in Large Language Models (LLMs) have significantly transformed the landscape of natural language processing. While LLMs have demonstrated outstanding performance in text-based tasks \cite{adeshola2024opportunities}, there is growing interest in extending their capabilities to handle other modalities, such as speech \cite{tang2023salmonn,li2024audio}. Typically, these speech LLMs integrate pre-trained speech encoders with text-based LLMs via projection layers that align feature dimensions across modalities\cite{wu2024prompting}. 

While LLMs are widely used in writing assessment \cite{mizumoto2023exploring, yancey2023rating}, their application to spoken language assessment is still emerging. Previous work, such as \cite{wang2025exploring}, explored text-only models like ChatGPT for pronunciation feedback. 
To overcome the limitations of text-only models, multimodal LLMs like GPT-4o \cite{hurst2024gpt} have been developed to process raw audio input. These models assess pronunciation across dimensions such as accuracy, fluency, prosody, and sentence completeness, and generate context-aware feedback. Recent studies \cite{wu2024prompting, wu2025integrating} introduced Audio-Text Prompt LLMs using wav2vec2 encoders and LLaMA2 decoders, enhancing performance by embedding mispronunciations into prompts. Models like Qwen-Audio \cite{chu2023qwen, chu2024qwen2}, extensively trained on both supervised and unsupervised speech data at a large scale, followed by instruction tuning, have shown good results in speech tasks such as ASR, translation, audio captioning, and spoken question answering. \cite{ma2025assessment} explored the potential of speech LLM for L2 oral proficiency assessment by evaluating zero-shot capabilities and proposing fine-tuning strategies, ultimately demonstrating that adapted models significantly outperform traditional baselines.

Building on recent advances in audio-text prompt-based language models, we examine the potential of the Qwen-Audio-Instruct model for spoken language assessment. We adopt the term multi-aspect to refer to the simultaneous evaluation of multiple dimensions of speech performance, i.e., accuracy, fluency, prosody, and sentence completeness, rather than focusing solely on phoneme-level correctness \cite{9746743}.  To the best of our knowledge, this is the first work that evaluates an instruction-tuned, speech LLM in a zero-shot setting for multi-aspect pronunciation assessment using rubric-based scoring. The model combines a powerful audio encoder with an instruction-tuned decoder, making it a strong candidate for the holistic evaluation of L2 speech. In this work, we explicitly ask the following research question: \textit{To what extent can the Qwen-Audio-Instruct model perform multi-aspect spoken language assessment, covering accuracy, fluency, prosody, and sentence completeness, and generate rubric-aligned scores in a zero-shot setting?}

\section{Methodology}

We evaluate the zero-shot performance of the Qwen2-Audio-7B-Instruct model\footnote{\tiny\url{https://huggingface.co/Qwen/Qwen2-Audio-7B-Instruct}} for multi-aspect spoken language assessment. This instruction-tuned, multimodal large language model (LLM) integrates a Whisper-based speech encoder and a transformer-based text decoder, enabling direct audio-text interaction without task-specific fine-tuning. We adopt a zero-shot setting to assess the model's generalization ability—specifically, its capacity to follow rubric-based instructions without prior adaptation. Zero-shot evaluation has emerged as a compelling paradigm for assessing the transfer capabilities of instruction-tuned models, particularly in low-resource or domain-agnostic scenarios \cite{wei2022emergent, sun2023evaluating}.

\subsection{Dataset}

We use the Speechocean762 dataset \cite{zhang2021speechocean762}, a publicly available corpus designed for L2 pronunciation assessment. It comprises 5,000 English utterances from 250 Mandarin-speaking learners, with balanced representation across children and adults. Each utterance is manually rated by five expert annotators at the sentence, word, and phoneme levels. For this study, we focus on the sentence-level annotations, which include four rubrics: \textbf{accuracy}, \textbf{fluency}, \textbf{prosody}, and \textbf{completeness}. These scores are provided on a 0–10 scale and are suitable for testing multi-aspect scoring in a zero-shot setting.

\subsection{Model Architecture}

The Qwen2-Audio model comprises two primary components:
\begin{itemize}
    \item \textbf{Audio Encoder:} Based on Whisper-large-v3 \cite{radford2023robust}, it converts raw waveforms into mel-spectrograms (16kHz) and produces downsampled audio embeddings aligned with the language model's input space.
    \item \textbf{Text Decoder:} Built on the Qwen-7B architecture \cite{bai2023qwen}, a decoder-only transformer trained on trillions of tokens. It receives the audio embeddings and instruction prompts to generate structured outputs. Instruction-following behavior is enhanced through supervised fine-tuning and Direct Preference Optimization (DPO).
\end{itemize}

\subsection{Prompting Strategy}




Each learner's utterance is paired with a single, multi-modal prompt containing both the raw audio and the expected reference text.\footnote{\tiny{\url{https://github.com/Aditya3107/sentence-scoring-qwen}}} 
This strategy allows the model to simultaneously process and assess pronunciation across rubrics, thereby reflecting a realistic, integrated evaluation scenario.
In particular, the model is instructed to evaluate the speaker’s pronunciation holistically using four sentence-level rubrics: \emph{accuracy}, \emph{fluency}, \emph{prosody}, and \emph{sentence completeness}. These rubrics are adapted from the original Speechocean762 guidelines \cite[Section~3.1, Table~1]{zhang2021speechocean762}, and refined using detailed descriptions to enhance the model's interpretability.

The prompt explicitly defines each rubric with a scoring scale from 0 to 10, along with criterion-referenced descriptions for each level. For instance, \emph{accuracy} is defined in terms of the number and severity of mispronunciations, while \emph{fluency} considers speech coherence, pause distribution, and hesitations. \emph{Prosody} focuses on the use of intonation, rhythm, and stress patterns to convey natural speech flow. \emph{Completeness} refers to the extent to which the entire target sentence was spoken, regardless of pronunciation correctness or intelligibility. The prompt is phrased to reflect a real-world assessment context by assuming the model plays the role of a pronunciation expert rating a non-native learner. An excerpt from the actual prompt provided to the model is shown below. The system is instructed as follows:

\begin{quote}
\small
\begin{lstlisting}
The speaker was supposed to say: "<target sentence>"
You are a pronunciation assessment expert. The speaker is a non-native English learner.
Listen carefully and provide the  sentence-level scores using these rubrics:
- Accuracy (0-10): 9-10 = Excellent pronunciation without mispronunciation to 0-2 = Unintelligible. 
- Fluency (0-10): 9-10 = Smooth and natural to 0-2 = Severely disfluent.
- Prosody (0-10): 9-10 = Good intonation and rhythm to 0-2 = Flat or stammered.
- Completeness (0-10): Whether the sentence was fully spoken.
Return your response in JSON format.
\end{lstlisting}
\end{quote}

The full prompt includes all four rubrics with five-point scoring criteria between 0-10 and instructs the model to return its ratings in a structured JSON format. An example target output might look like: \texttt{\{"Accuracy": 4, "Fluency": 3, "Prosody": 3, "Completeness": 5\}}.

To extract the final scores from the model's output, we employ a regular-expression-based parser that scans for dictionary-style text fragments in the model response and maps them to the four rubric categories. This process ensures robustness against minor formatting deviations while preserving interpretable score extraction. Invalid or missing fields are marked as \texttt{None} and excluded from the final analysis. Nonetheless, all utterances yielded responses, and no model output was discarded.

\subsection{Evaluation Protocol}

The model's outputs are evaluated against the human-annotated sentence-level scores from the Speechocean762 dataset, which are treated as ground truth (GT). By using the same rubric definitions in both the prompt and the GT, we ensure a fair alignment between model-generated and human-provided evaluations. All 5,000 utterances are used to benchmark zero-shot performance across the four scoring dimensions.

\section{Results}
We evaluated the Qwen-Audio-Instruct model's ability to predict sentence-level pronunciation scores across four rubrics \cite{zhang2021speechocean762}. The evaluation was conducted on a dataset of 5,000 utterances from the SpeechOcean762 corpus. We focus on analyzing the model’s scoring behavior, match rates, correlation with human ratings, and sensitivity to speech quality.

\begin{figure*}[ht!]
    \centering
    \includegraphics[width=\linewidth]{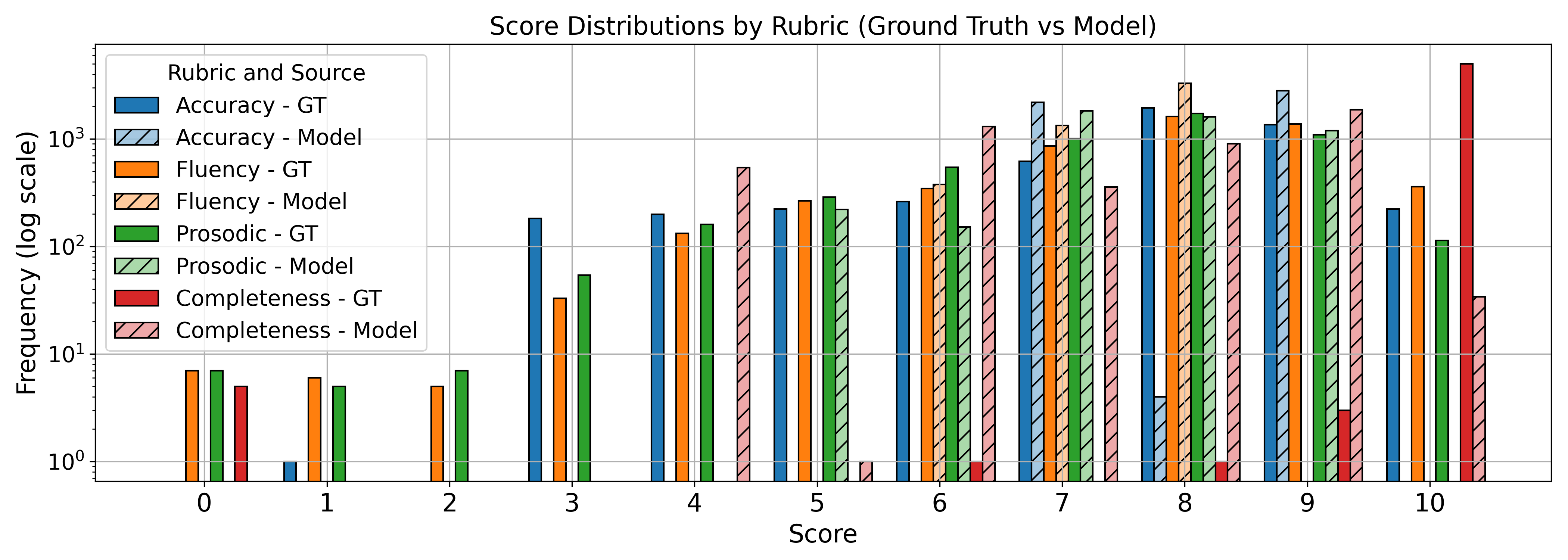}
    \caption{Score distributions for GT and model predictions across all rubrics. GT bars are solid and opaque, while model predictions use lighter colors with striped hatch patterns. Each rubric is shown using a distinct color with grouped bars.}
    \label{fig:score_distribution}
\end{figure*}

To understand how closely the model predictions aligned with human scores, we compared the score distributions using histograms. Figure \ref{fig:score_distribution} shows that the model exhibited a strong central bias, particularly assigning scores between 7 and 9, while largely avoiding lower scores across all rubrics. This pattern was most pronounced in accuracy and fluency, with minimal model predictions below a score of 7. In contrast, prosody stood out as the rubric where the model distribution more closely mirrored the human annotations.

Quantitatively, as shown in Table \ref{tab:score_match}, exact matches between the model and GT scores were observed in 23.3\% of cases for accuracy, 26.1\% for fluency, 25.4\% for prosody, and only 0.7\% for completeness. While these exact match rates are relatively low, around 20–25\% for most rubrics, they establish a strict baseline for alignment, indicating how often the model replicates human scores without calibration. 

With a relaxed margin of $\pm$1 point, agreement increased to 68.1\% for accuracy, 69.0\% for fluency, 67.3\% for prosody, and 37.8\% for completeness. Under a broader $\pm$2 margin, alignment further improved to 87.4\%, 89.5\%, 87.5\%, and 55.9\%, respectively. These results suggest reasonable agreement with human judgments, especially in the mid-to-high score range.

However, it is important to note that a $\pm$2 margin on a 10-point scale spans 40\% of the total range, reducing scoring granularity. Thus, while higher agreement at this margin reflects general alignment, it should be interpreted with caution.

\begin{table}[ht!]
\centering
\small
\caption{Match rates between model predictions and human scores, reported as exact and within $\pm$1 and $\pm$2 points.}
\label{tab:score_match}
\begin{tabular}{lccc}
\toprule
\textbf{Rubric} & \textbf{Exact Match (\%)} & \textbf{$\pm$1 (\%)} & \textbf{$\pm$2 (\%)} \\
\midrule
Accuracy     & 23.3 & 68.1 & 87.4 \\
Fluency      & 26.1 & 69.0 & 89.5 \\
Prosody      & 25.4 & 67.3 & 87.5 \\
Completeness &  0.7 & 37.8 & 55.9 \\
\bottomrule
\end{tabular}
\end{table}

\subsection{Score Correlations}
\label{distribution_agreement}

To further quantify the model’s alignment with human ratings, we calculated Pearson correlation coefficients (PCC) between GT and model scores. In addition to raw score correlations, we computed PCCs using binary relaxed-match indicators, denoting whether the model predictions fall within $\pm$1 or $\pm$2 points of the GT. This tolerance-based analysis provides a more flexible evaluation of model-human agreement by considering near matches as acceptable outcomes. As shown in Table~\ref{tab:combined_pcc}, direct score alignment resulted in relatively low correlations, with accuracy achieving the highest ($r = 0.140$), followed by prosody ($r = 0.206$), while fluency and completeness remained weak ($r = 0.053$ and $r = -0.021$, respectively). In contrast, the tolerance-based PCCs revealed stronger alignment across most rubrics, particularly for accuracy ($r = 0.466$ at $\pm$1 and $r = 0.671$ at $\pm$2) and prosody ($r = 0.365$ at $\pm$1 and $r = 0.422$ at $\pm$2). Fluency also showed moderate improvement ($r = 0.161$ at $\pm$1 and $r = 0.373$ at $\pm$2), while completeness continued to exhibit poor alignment even under relaxed conditions ($r = 0.031$ at $\pm$1 and $r = 0.037$ at $\pm$2).

\begin{table}[ht!]
\centering
\small
\caption{PCCs between GT scores and model predictions, including tolerance-based matches within $\pm$1 and $\pm$2 points.}
\label{tab:combined_pcc}
\begin{tabular}{ccc}
\toprule
\textbf{Rubric} & \textbf{GT vs Model} & \textbf{±1 / ±2 Match} \\
\midrule
Accuracy     & 0.206 & 0.466 / 0.671 \\
Fluency      & 0.053 & 0.161 / 0.373 \\
Prosody      & 0.140 & 0.365 / 0.422 \\
Completeness & -0.021 & 0.031 / 0.037 \\
\bottomrule
\end{tabular}
\end{table}



\subsection{Performance on Low-Quality Utterances}
\label{low_quality}

The Qwen-audio Instruct model consistently failed to penalize poor pronunciation adequately. For utterances with GT accuracy scores less than or equal to 6 (\(n = 863\)), the model did not assign correspondingly low scores (i.e., \(\leq 6\)) in any instance. A similar pattern of overestimation was evident in the fluency and prosody dimensions, where the model failed to assign low scores in 95.7\% and 89.3\% of the respective low-GT cases. These results indicate a systematic bias toward overestimation, particularly in the evaluation of lower-quality utterances.

\subsection{Sensitivity to Ground Truth Quality}

To further assess the model's sensitivity to speech quality, we compared its average predicted scores based on GT score thresholds. When GT scores were \(\geq7\), model predictions were consistently higher across all rubrics. Specifically, average model scores increased from 7.71 to 8.21 for accuracy, from 7.55 to 7.59 for fluency, from 7.44 to 7.73 for prosody, and from 6.95 to 7.45 for completeness. While these differences confirm some responsiveness to GT scores, the relatively narrow margins, especially in fluency, suggest limited granularity in the model's evaluative capability.

\section{Discussion}
This study examined the extent to which the Qwen-Audio-Instruct model can perform multi-aspect spoken language assessment, covering accuracy, fluency, prosody, and sentence completeness, and generate rubric-aligned scores in a zero-shot setting without any task-specific fine-tuning. Our findings reveal a nuanced picture of both the capabilities and current limitations of instruction-tuned speech LLMs applied to spoken language evaluation.

Despite not being explicitly trained or fine-tuned for speech assessment or rubric scoring, Qwen-Audio-Instruct approximated human ratings across multiple dimensions. When evaluated with a relaxed margin of $\pm2$, the model achieved over 85\% match rates for accuracy, fluency, and prosody (Table~\ref{tab:score_match}), especially in high-quality speech. This suggests that instruction-tuned LLMs can perform coarse-grained, multi-faceted evaluations without adaptation—useful for rapid prototyping or low-resource deployments.

When evaluated with a relaxed margin of error, the model’s predictions aligned reasonably well with human ratings, especially in higher-quality speech ranges (see Section~\ref{distribution_agreement}). This indicates that instruction-tuned LLMs can leverage their broad language understanding and acoustic representations to perform coarse-grained, multi-faceted speech evaluations out of the box. Such zero-shot generalization is particularly promising for low-resource scenarios, rapid prototyping, and initial feasibility assessments where labeled data and fine-tuning resources are limited. 

Among the four rubrics, prosody appeared visually closer to the ground truth in terms of score spread (Figure~\ref{fig:score_distribution}), with a similar peak frequency and overall distribution shape. This suggests that the model may capture suprasegmental features, such as intonation and rhythm, more effectively than segmental aspects like phoneme accuracy. This could indicate a relatively stronger internal representation of prosodic patterns, even in a zero-shot setup. However, the model’s prosody predictions lacked low-end coverage, as no scores below 5 were assigned, limiting its ability to reflect the full range of human annotations. Furthermore, this visual similarity did not correspond to the highest quantitative agreement: accuracy achieved the strongest correlations with human ratings under relaxed matching conditions (Table~\ref{tab:combined_pcc}), while fluency showed the highest match rates overall (Table~\ref{tab:score_match}).

In particular, the sentence completeness rubric presents a unique challenge. As shown in the histogram in Figure \ref{fig:score_distribution}, human ratings for completeness were overwhelmingly biased toward the maximum score of 10, even in cases where they included unknown words or diverged from the expected sentence. This likely reflects a lack of consistency or clear guidance in the human annotation process. Notably, the completeness rubric in the Speechocean762 dataset is not defined with the same level of detail as the other dimensions; it is loosely described as the proportion of words in the target text that are pronounced. In line with this, the model was instructed to assign a score for completeness without an explicit rubric definition. This ambiguity may have contributed to the significant discrepancy between human and model scores for completeness, as observed in both the distribution and correlation results (see Table~\ref{tab:combined_pcc}). Addressing this rubric-specific gap is critical for improving both annotation quality and model alignment.

However, the model consistently overpredicted scores for lower-quality or error-prone utterances, revealing important challenges. For example, among 863 utterances with ground truth accuracy scores  $\leq 6$, the model did not assign a single score in this range. These findings point to a systematic overestimation bias, particularly in handling lower-quality utterances. Several factors likely contribute to this tendency. First, the model’s training objectives, which emphasize helpfulness and politeness, may bias it against harsh or low-scoring responses unless specifically prompted. Second, a probable scarcity of negative or poorly pronounced examples during pretraining could skew the output toward mid-to-high scores. Third, prompt formulations may insufficiently highlight the importance of differentiating low-end performance, causing the model to default to “safe” central values. Moreover, the model’s reliance on abstracted high-level acoustic embeddings might obscure subtle phonetic and prosodic errors, limiting precision in error detection. The low performance on the completeness rubric suggests that content coverage and discourse-level evaluation remain challenging without detailed alignments or ground-truth references.

\section{Conclusion}

In summary, the Qwen-Audio-Instruct model shows strong potential for scalable and flexible multi-aspect speech evaluation in zero-shot settings. However, key challenges remain in achieving fine-grained, interpretable, and calibrated scoring, especially for lower-quality speech and nuanced phonetic deviations. Addressing these issues will require domain-specific fine-tuning, better prompt strategies for low-end scoring, and integration of phonetic-level representations or alignment mechanisms.

Pursuing these directions may position instruction-tuned speech LLMs as powerful tools for CAPT and language learning, bridging the gap between model adaptability and the precision needed for reliable language proficiency evaluation.

\section{Acknowledgements}
This publication is part of the project Responsible AI for Voice Diagnostics (RAIVD) with file number NGF.1607.22.013 of the research programme NGF AiNed Fellowship Grants which is financed by the Dutch Research Council (NWO).

\bibliographystyle{IEEEtran}
\bibliography{mybib}

\end{document}